\documentclass[11pt]{article}
  
\usepackage{amsmath,amssymb}

\usepackage{epsfig}  
\usepackage{graphicx}               % Standard graphics package  
\usepackage{url}
\usepackage{hyperref}

\usepackage{pstricks}
\usepackage{color}
\usepackage{cite}
 
\newcommand{\nc}{\newcommand}  

\nc{\beq}{\begin{equation}}  
\nc{\eeq}{\end{equation}}  
\nc{\beqa}{\begin{eqnarray}}  
\nc{\eeqa}{\end{eqnarray}}  
\nc{\bea}{\begin{eqnarray}}  
\nc{\eea}{\end{eqnarray}}  
\nc{\ra}{\rightarrow}  
\nc{\lsim}{\begin{array}{c}\,\sim\vspace{-21pt}\\< \end{array}}  
\nc{\gsim}{\begin{array}{c}\sim\vspace{-21pt}\\> \end{array}}  
\nc{\Tr}{{\rm Tr}}
\nc{\slsh}{\slash\hspace*{-0.22cm}}

\def\be{\begin{equation}}
\def\ee{\end{equation}}
\def\bea{\begin{eqnarray}}
\def\eea{\end{eqnarray}}
\def\bit{\begin{itemize}}
\def\eit{\end{itemize}}

\newcommand{\vev}[1]{ \left\langle {#1} \right\rangle }

\newcommand{\calH}{{\cal H}}

\def\to{\rightarrow}

\newcommand{\ov}[1]{\overline{#1}}

\title{  
\vspace*{-2.3cm}  
\begin{flushright}  
\normalsize{  
  }  
\end{flushright}  
\vspace{1.5cm}  
\Large  
\textbf{
A Holographic Model of Heavy-light Mesons
 \\
}\vspace*{1.0cm}   
}

\author{Yang Bai$\,^{a}$ and Hsin-Chia Cheng$\,^{b}$\vspace{5mm}
\\
$^{a}$ \normalsize\emph{Department of Physics, University of Wisconsin, Madison, WI 53706, USA}
\vspace{1mm} \\
$^{b}$ \normalsize\emph{Department of Physics, University of California, Davis, CA 95616, USA}
}

\date{}

\begin{document}  
\setcounter{page}{0}  
\maketitle  

\vspace*{1cm}  
\begin{abstract} 
We construct a holographic model of heavy-light mesons by extending the AdS/QCD to incorporate the behavior of the heavy quark limit. In that limit, the QCD dynamics is governed by the light quark and the heavy quark simply plays the role of a static color source. The heavy quark spin symmetry can be treated as a global symmetry in the AdS bulk. As a consequence, the heavy-light mesons are mapped to ``fermions'' in the AdS theory. The light flavor chiral symmetry is naturally built in by this construction, and its breaking produces the splitting of the parity-doubled heavy-light meson states. The scaling dependences of physical quantities on the heavy quark mass in the heavy quark effective theory are reproduced. The mass spectra and decay constants of the $B$ and $D$ mesons can be well fit by suitable choices of model parameters.  The couplings between the heavy-light mesons and the pions are also calculated. The holographic model may capture the essence of the long distance effects of QCD and can serve as a useful tool for studying the non-perturbative hadronic matrix elements involving heavy-light mesons.
\end{abstract}  
  
\thispagestyle{empty}  
\newpage  
  
\setcounter{page}{1}

\baselineskip18pt   

\vspace{-3cm}

%-----------------------------------------------------------------------------
\section{Introduction}
\label{sec:introdction}
%-----------------------------------------------------------------------------

The bottom-up AdS/QCD~\cite{Erlich:2005qh,DaRold:2005zs,Hirn:2005nr} attempts to approximate the low-energy Quantum Chromodynamics (QCD) by a five-dimensional (5D) theory living in a slice of anti-de Sitter (AdS) space using the AdS/CFT correspondence~\cite{Maldacena:1997re,Gubser:1998bc,Witten:1998qj}. Even though it is not derived from the first principle and the real QCD is neither conformal nor possessing large number of colors ($N_c$), it has worked reasonably well in describing the low-energy mesons made of light quarks.  Many features of the low-energy QCD, such as Vector Meson Dominance~\cite{Sakurai:1960ju} and Hidden Local Symmetry~\cite{Bando:1984ej}, are built in the AdS/QCD. Its success may be viewed as that it captures some essence of the strong dynamics of QCD.

The simplest version of AdS/QCD describes the $J^{PC} = 1^{--}$ vector mesons, $1^{++}$ axial vector mesons, and the $J^P= 0^-$ Nambu-Goldstone bosons (pions) associated with the chiral symmetry $SU(N_f)_L\times SU(N_f)_R$ for $N_f$ light quarks. The action is given by
\begin{equation}
{\cal S} = \int d^5 x M_5 \sqrt{g}\, \Tr \left[  -\frac{1}{2} (L_{MN}L^{MN} +R_{MN}R^{MN})  + |D_M \Sigma|^2 - M_\Sigma^2 |\Sigma|^2\right] ,
\end{equation} 
with the AdS metric
\begin{equation}
ds^2 = \frac{R^2}{z^2} (\eta_{\mu\nu} d x^\mu d x^\nu  -d z^2),
\end{equation}
between the UV-boundary ($z= \epsilon$) and the IR-boundary ($z=L_1$), where $R$ is the  AdS curvature radius. The $L_{MN}$ and $R_{MN}$ are the field strength tensors of the $SU(N_f)_L$ and $SU(N_f)_R$ gauge symmetry in the bulk which are associated with the corresponding $SU(N_f)_L$ and $SU(N_f)_R$ current operators of QCD. The scalar field $\Sigma$ transforms as $({N_f}, \bar{N}_f)$ which corresponds to the $\bar{q}_R q_L$ operator in QCD. $D_M$ is the gauge covariant derivative, $D_M \Sigma = \partial_M \Sigma+i L_{M} \Sigma -i \Sigma R_{M}$.  $M_5$ is related to the 5D gauge coupling, $M_5 =1/g_5^2$, and is taken as the 5D fundamental scale. The solution of $\Sigma$ in the bulk takes the form:
\begin{equation}
\vev{\Sigma (z)} =  \left[ \frac{M_q}{R}  z\left(\frac{z}{R}\right) ^{3-\Delta_\Sigma} + \frac{\xi}{R L_1^3} z^3 \left(\frac{z}{R}\right)^{\Delta_\Sigma-3} \right] \mathbb{I}_{N_f},
\end{equation}
where $\Delta_\Sigma$ is the scaling dimension of the $\bar{q} q$ operator and is related to the $\Sigma$ field bulk mass by $\Delta_\Sigma (4-\Delta_\Sigma) = M_\Sigma^2 R^2$. The first term is associated with the light quark mass $M_q$ which corresponds to an explicit chiral (and conformal) symmetry breaking effect, and the coefficient of the second term is related to the vacuum expectation value (VEV) of the $\bar{q} q$ operator, which spontaneously breaks the chiral (and conformal) symmetry. The model has few parameters and can be used to fit a wide range of light meson data. The number of parameters can be further reduced if one matches them to the perturbative QCD results as was done in the original AdS/QCD papers~\cite{Erlich:2005qh,DaRold:2005zs,Hirn:2005nr}. There, the scaling dimension was taken to be the na\"ive dimension of the $\bar{q} q$ operator, $\Delta_\Sigma =3$. As a consequence, the predictions depend on the combination $M_5 R$ but not on $M_5$ or $R$ separately. If one further matches the two-point function in the UV to the perturbative QCD result, one finds
\begin{equation}
M_5 R = \frac{N_c}{12 \pi^2}\,.
\end{equation}
After taking the position of the UV-boundary $\epsilon$ to 0, the predictions of this simplest model only depend on three parameters: the light quark mass $M_q$, the position of the IR-boundary $L_1$ which corresponds to the confinement scale, and $\xi$ which represents the ratio of chiral symmetry breaking and the confinement scale, both of which are related to a common QCD scale $\Lambda_{\rm QCD}$. They can be chosen to fit the light meson spectrum. Specifically, to fit the $\rho$ and $a_1$ masses it was found that $L_1^{-1} \approx 320$~MeV and $\xi \approx 4$, and $M_q$ can be obtained by fitting the $\pi$ mass~\cite{DaRold:2005zs}. The theory can then be used to calculate a variety of low-energy quantities, including the mass spectrum of the excited meson states, decay constants, couplings among meson states, and coefficients of the chiral Lagrangian. A reasonable agreement with the experimental measurements has been found for the ground states. The spectrum of the  higher excited meson states does not follow the Regge trajectory in this simple hard-wall model where there is a sharp IR cutoff at $z=L_1$, but it can be improved by introducing a soft-wall potential in the bulk~\cite{Karch:2006pv}. Given the simplicity of the model and its crude approximation to the real QCD, the extent of the agreement with the real QCD data is quite impressive.  

The success faces challenges when one tries to include the $1^{+-}$ $h_1/b_1$-like mesons. They are created by the dimension-3 tensor operator, $\bar{q} \sigma^{\mu\nu} T^a q$ which are associated with a two-form field in the AdS bulk~\cite{Cappiello:2010tu,Domokos:2011dn,Alvares:2011wb}. In particular, if one also requires the new parameters related to the two-form field sector to be matched to the perturbative QCD values, the predictions of the AdS/QCD do not match well with the actual data and even the success of $1^{--}$ meson sector is ruined due to mixing of the vector and tensor operators~\cite{Alvares:2011wb}. However, it was argued in Ref.~\cite{Domokos:2012da} that there is no reason to insist that the parameters in AdS/QCD should be matched to the perturbative QCD values. The two theories have different UV limits and the renormalization group (RG) running in the real QCD can change the parameters in the IR. Therefore, it was advocated in Ref.~\cite{Domokos:2012da} that the parameters other than those protected by symmetries should be treated as free parameters to be fit from the experimental data. It turns out that the best-fit values for the parameters in the original AdS/QCD are close to the old values matched perturbative QCD, while the new parameters involving the two-form field need to take different values~\cite{Domokos:2012da}. In that case, at least the success of the original hard-wall AdS/QCD is preserved though the predictions of the $b_1$ sector are not as good. 

Because AdS/QCD and the real QCD have different UV limits, one should not expect AdS/QCD to be a good model for QCD at high energies far above $\Lambda_{\rm QCD}$. As shown in Ref.~\cite{Csaki:2008dt}, the event shape of the AdS/QCD in high energy collisions is more spherical with high multiplicities, unlike the jetty structure in the real QCD. Indeed, at high energies the QCD coupling is perturbative and there is no need to choose a dual theory where the coupling is strong and perform calculations there. For the same reason, AdS/QCD may not be a good approximation when applied to heavy quarkonium states~\cite{Erdmenger:2007vj,Erdmenger:2007cm,White:2007tu,Kim:2007rt}. An interesting question is whether AdS/QCD can provide a good approximation to QCD bound states made of both heavy and light quarks, in particular, the heavy-light mesons such as $B$ and $D$ mesons. In the heavy quark limit, the heavy quark in a heavy-light meson just plays the role of a static color source and the dynamics is governed by the light quark. From this point of view, one might expect that the success of the AdS/QCD for the light mesons could be carried over to the heavy-light meson system. There have been studies of AdS/QCD for the heavy-light mesons in the top-down approach with string and brane constructions as well as the light-front holography~\cite{Erdmenger:2006bg,Erdmenger:2007vj,Erdmenger:2007cm,Herzog:2008bp,Branz:2010ub}. In this paper we follow the bottom-up approach of Ref.~\cite{Erlich:2005qh,DaRold:2005zs} and extend it to the heavy-light meson system. We try to fit the real experimental or lattice $B$ and $D$ meson data and hope that such a model can reproduce the qualitative feature of the non-perturbative aspects of the heavy-light mesons.

In the heavy quark limit, the heavy-light mesons exhibit the heavy quark spin symmetry $SU(2)_h \times SU(2)_l$. The scalar and vector mesons related by the spin symmetry become degenerate in that limit. It is convenient and commonly done in the heavy quark effective theory (HQET) to express them as a bi-spinor field where the spin symmetry can be made manifest. (For a review of the HQET, please see Ref.~\cite{Manohar:2000dt}.) Since the heavy quark is static, its fermionic nature plays no role other than providing the multiplicity of the spin states. One might as well treat the heavy quark as a boson and the heavy quark spin symmetry as a global symmetry. The light quark component, on the other hand, participates in the strong dynamics which may be modeled by AdS/QCD. This suggests that in AdS/QCD, the heavy-light mesons should be mapped to ``fermions'' in the AdS bulk, with the heavy quark spin symmetry treated as a bulk flavor symmetry. We show that in such a setup, which we dub AdS-HQET, many heavy-light meson data can be described in the AdS/QCD model with suitable parameters. It may provide qualitative insights of nonperturbative effects of processes involving heavy-light mesons. Since AdS/QCD is at best a crude approximation for the real QCD, we only focus on the leading effects in the heavy quark limit. Effects suppressed by the heavy quark mass such as the mass splitting between the spin-0 and spin-1 mesons from the hyperfine interaction will not be considered in this paper.

This paper is organized as follows. In Sec.~\ref{sec:match} we review the HQET formalism for heavy-light mesons and set up our notations and convention. We then derive the fermionic Lagrangian in the static heavy quark limit, which serves as the starting point to construct the holographic AdS-HQET model. In Sec.~\ref{sec:5D-lagrangian}, we incorporate the heavy-light mesons into AdS/QCD in the chiral limit as an illustration of the construction and calculation techniques. In Sec.~\ref{sec:parity-double} the chiral symmetry breaking and the splitting between the parity doublets of the heavy-light mesons are introduced. We perform fits of the spectrum and decay constants to the experimental and lattice data to determine the model parameters. We also calculate the coupling of the heavy-light mesons to the pions. The future applications of the AdS-HQET model, such as computations of weak-interaction processes involving heavy-light mesons, are discussed in Sec.~\ref{sec:conclusions}.

%-----------------------------------------------------------------------------
\section{Effective Lagrangian for Heavy-light Mesons}
\label{sec:match}
%-----------------------------------------------------------------------------
In this section we review the effective Lagrangian for the heavy-light mesons and show that they can be put in a form of the fermion Lagrangian which will be our starting point to incorporate them into AdS/QCD. We follow the notation of HQET in Ref.~\cite{Bardeen:2003kt} by Bardeen, Eichten and Hill (BEH), in which the spin-zero and spin-one mesons are combined to be written as a velocity-dependent bi-spinor field
\beqa
{\cal H}_v = ( i \gamma_5 H_v + \gamma_\mu H_v^\mu) \left(\frac{1+\slsh{v}}{2}\right) \,.
\eeqa
Here, $H_v$ ($0^-$) and $H_v^\mu$ ($1^-$) represent spin-zero and spin-one mesons, respectively, and the velocity-dependent field is related to the original field by
\beqa
H_v = \sqrt{M} e^{i M v\cdot x} H,
\eeqa
where $M$ represents the heavy quark mass.\footnote{This definition differs from that of BEH by $\sqrt{2}$ for later convenience.}
We have chosen the first index in ${\cal H}$ to be the light quark spinor index and the second index to be the heavy quark spinor index. The field ${\cal H}_v$ satisfies ${\cal H}_v\,\slsh v = {\cal H}_v$ and $\slsh v\, {\cal H}_v = - {\cal H}_v$ using the relation $v_\mu H^\mu=0$ for physical spin-one particles. It was shown in the Appendix of Ref.~\cite{Bardeen:2003kt} that to order $1/M$ the free Lagrangian of ${\cal H}_v$ can be written as
\beqa
{\cal L}_0= -i \,\Tr ( \overline{\cal H}_v \, v \cdot \partial {\cal H}_v) + \delta M\, \Tr (\overline{\cal H}_v {\cal H}_v) \, ,
\label{eq:HQETL}
\eeqa
where $\delta M\ll M$ represents the difference between the meson mass and the heavy quark mass. The division between $M$ and $\delta M$ is somewhat arbitrary and for convenience we can ``gauge away'' $\delta M$~\cite{Bardeen:2003kt}.
Similarly, we have the bi-spinor ${\cal H}^\prime_v$ for parity-even states constructed from $H^\prime$ ($0^+$) and $H^{\prime\, \mu}$ ($1^+$). Combining ${\cal H}_v$ and ${\cal H}^\prime_v$ we can form linear representations of the light flavor chiral symmetry $SU(N_f)_L\times SU(N_f)_R$, 
\beqa
{\cal H}_{L\,v} = \frac{1}{\sqrt{2}} ( {\cal H}^\prime_v -  \,{\cal H}_v) \,, \qquad 
{\cal H}_{R\,v} = \frac{1}{\sqrt{2}} ( {\cal H}_v +  \,{\cal H}^\prime_v) \,,
\label{eq:HHprime}
\eeqa
with ${\cal H}_{L\,v}$ transforming as $(N_f, { 1})$ and ${\cal H}_{R\,v}$ transforming as $({ 1}, {N_f})$ under the chiral symmetry. It was argued that in the chiral symmetry limit, ${\cal H}_{L\,v}$ and ${\cal H}_{R\,v}$ are degenerate and form a parity-doublet~\cite{Bardeen:1993ae,Bardeen:2003kt,Nowak:1992um,Nowak:2003ra}.

The Lagrangian in Eq.~(\ref{eq:HQETL}) can also be written equivalently as
\beqa
{\cal L} = \mbox{Tr} \left( \ov{\calH}\, i\slsh \partial\, \calH   \right) \,+\, M\,\mbox{Tr} \left(  \ov{\calH}\, \calH \right) \,,
\label{eq:fermionL}
\eeqa
if we define ${\cal H} \equiv e^{- i M v\cdot x}  {\cal H}_v$. It looks like a fermion Lagrangian except that the adjoint of the bi-spinor is defined with $\gamma^0$ multiplying on both spinor indices, $\overline{\cal H} = \gamma^0 {\cal H}^\dagger \gamma^0$. In the Pauli-Dirac representation, $\gamma^0$ is given by
\beqa
\gamma^0 = \begin{pmatrix} \mathbb{I}_2 & 0  \\ 0 & -\mathbb{I}_2  \end{pmatrix}, \quad \mbox{where $\mathbb{I}_2$ is the $2\times 2$ unit matrix.}
\eeqa
If we treat the heavy quark spinor index as a flavor index, this Lagrangian simply describe four species of fermions with the last two fermions having the opposite sign in the Lagrangian. Since the bi-spinor fields always appear in pairs, we can redefine the field to absorb the minus sign in the path integral, i.e., treating ${\cal H}^\dagger$ and ${\cal H}$ as independent fields and absorb the $\gamma^0$ multiplied on the heavy quark spinor index into ${\cal H}^\dagger$, then it takes the standard form of the fermion Lagrangian. The reason that we can describe the heavy-light mesons by fermion fields simply reflects the fact that the heavy quark just plays the role of a static color source and whether it is a fermion or a boson does not affect the dynamics, as long as we do not include heavy quark loops in the calculation. 

In the fermionic theory, we introduce a global flavor symmetry $SU(2)_f$ to match the heavy spin symmetry $SU(2)_h$. Specifically, we consider two copies of four Weyl fermions, $\psi^{k}_{1,L}$, $\psi^{k}_{1,R}$, $\psi^{k}_{2,L}$, $\psi^{k}_{2,R}$, where  ``$k=1, 2$" is the flavor index which represents the degrees of freedom coming from the heavy quark. Each Weyl fermion of course has a Lorentz spinor index ``$s=1, 2$" which corresponds to the spin degrees of freedom of the light quark. These Weyl fermions can be put into a $4\times 4$ matrix form:
\beq
{\cal H}^{\rm Weyl} = \begin{pmatrix} \psi_{1,L} & \psi_{2,L} \\ -\psi_{2,R} & - \psi_{1,R} \end{pmatrix},
\eeq
where the minus signs are just a convention. Just like the bi-spinor in the HQET, the first index of ${\cal H}^{\rm Weyl}$ is spinor index of the light quark (except that it is in the Weyl representation), while the second index corresponds to the global flavor symmetry which is matched to the heavy spin symmetry in the HQET. 
If we identify ${\cal H}^{\rm Weyl}$ with ${\cal H}$ and expand the Lagrangian of Eq.~(\ref{eq:fermionL}) (with only $\gamma^0$ on the light quark spinor side in the adjoint) in terms of the Weyl fermion components, the kinetic term and mass term are given by
\beqa
\mbox{Tr} ( \ov{\cal H}^{\rm Weyl}\, i \slsh \partial \, {\cal H}^{\rm Weyl} )  &=& \ov{\psi}_{1,L} i \bar{\sigma}^\mu \partial_\mu \psi_{1,L} + \ov{\psi}_{1,R} i \sigma^\mu \partial_\mu \psi_{1,R} + \ov{\psi}_{2,L} i \bar{\sigma}^\mu \partial_\mu \psi_{2,L} + \ov{\psi}_{2,R} i \sigma^\mu \partial_\mu \psi_{2,R}\,,
\label{eq:kineticterms}
\\
M\,\mbox{Tr}  (\ov{{\cal H}}^{\rm Weyl} \,{\cal H}^{\rm Weyl} ) &=& -M \left( \ov{\psi}_{1,L} \psi_{2,R} + \ov{\psi}_{2,R} \psi_{1,L} + \ov{\psi}_{2,L} \psi_{1,R} + \ov{\psi}_{1,R} \psi_{2,L} \right)\,,
\label{eq:massterms}
\eeqa
where the $SU(2)_f$ flavor indices are implicitly summed over. We see that it is indeed a standard Lagrangian describing four massive Dirac fermions.  

To match to the meson fields, it is more convenient to transform the fermions from the Weyl representation to the Pauli-Dirac representation using the transformation relation in Appendix~\ref{sec:weyldirac},\footnote{Here the transformation between the Weyl and Pauli-Dirac (PD) representations acts on the light spinor index only. For the heavy quark spinor index, the Pauli-Dirac representation is always used.}
\beqa
{\cal H}^{\rm PD} =\frac{1}{\sqrt{2}} \begin{bmatrix} \psi_{1,L}-\psi_{2,R} & \psi_{2,L}-\psi_{1,R} \\ -(\psi_{1,L}+\psi_{2,R}) & - (\psi_{2,L}+\psi_{1,R} )\end{bmatrix}.
\label{eq:bi-spinor-PD}
\eeqa
On the other hand, in the rest frame of the heavy quark, $v_\mu \rightarrow (1, 0, 0, 0)$, the projection operator $(1 + \slsh v)/2$ in the Pauli-Dirac representation takes the form
\beqa
\frac{1 +\slsh v}{2}  \rightarrow \frac{1 + \gamma^0}{2} = \begin{pmatrix} \mathbb{I}_2 & 0 \\ 0 & 0 \end{pmatrix},
\eeqa
and in terms of the spin-0 and spin-1 meson fields, ${\cal H}$ can be written as
\beqa
{\cal H} = \sqrt{M} (i \gamma_5 H + \gamma^\mu H^\mu) \frac{1 + \slsh v}{2} \rightarrow \sqrt{M} \begin{pmatrix} 0 & 0 \\ - \sigma^j H_j + i \mathbb{I}_2 H & 0 \end{pmatrix}.
\label{eq:mesons}
\eeqa
Comparing Eq.~(\ref{eq:mesons}) and (\ref{eq:bi-spinor-PD}) and matching the heavy-light mesons to the chiral fermions,  
we have the following dictionary:
\beqa
&&\psi_{1,L} + \psi_{2,R}  = \sqrt{2M}\,( \sigma^j H_j - i \,\mathbb{I}_2 H ) \,
, \\
\mbox{or}
&& H = \frac{i}{2\sqrt{2M}}\, \mbox{Tr} ( \psi_{1,L} + \psi_{2,R} ) \,, \quad \quad H^j = \frac{1}{2\sqrt{2M}} \mbox{Tr}\left[ \sigma^j (\psi_{1,L} + \psi_{2,R} ) \right] \,.
\label{eq:match-dictionary}
\eeqa
The number of degrees of freedom in the Weyl fermion combination, $\psi_{1,L}^{k} + \psi_{2,R}^{k}$,  are $2\times 2 =4$, which matches to that of one spin-zero meson $H$ plus one physical spin-one meson $H^j$.

%-----------------------------------------------------------------------------
\section{The AdS/QCD Model for Heavy-light Mesons in the Chiral Limit} 
\label{sec:5D-lagrangian}
%-----------------------------------------------------------------------------

We are now ready to write down the 5D AdS/QCD model for the heavy-light mesons. For simplicity we first consider the chiral limit and focus on the ${\cal H}_L$ sector. The effects of chiral symmetry breaking will be studied in the next section. The formalism developed in the previous section suggests that the heavy-light mesons should be represented by fermions in AdS/QCD. To include the heavy quark spin symmetry, we introduce two pairs of Dirac fermions in the AdS bulk,
\beqa
\Psi^k_1(x, z) = \left(\begin{array}{c}
\Psi^k_{1, L} (x, z)\\
\Psi^k_{1, R} (x, z)
\end{array}
\right)
\,,\qquad \mbox{and} \qquad
\Psi^k_2 (x, z)= \left(\begin{array}{c}
\Psi^k_{2, L} (x, z) \\
\Psi^k_{2, R} (x, z)
\end{array}
\right) \,,
\eeqa
where $k=1,2$ corresponds the heavy quark spin degree of freedom. For notational simplicity, the $k$ index will be suppressed in the rest of the paper.
The quadratic action for these fermions in the 5D AdS space between the UV cutoff $z=\epsilon$ and IR cutoff $z=L_1$ is given by
\beqa
{\cal S}_{5D} &\supset& M_5 \int d^5 x \left(\frac{R}{z} \right)^4 \left[
i \bar{\Psi}_{1, L} \bar{\sigma}^\mu \partial_\mu \Psi_{1, L} + i \bar{\Psi}_{1, R} \sigma^\mu \partial_\mu \Psi_{1, R}
- \frac{1}{2} ( \bar{\Psi}_{1, R} \overleftrightarrow{\partial_z} \Psi_{1, L} - \bar{\Psi}_{1, L} \overleftrightarrow{\partial_z} \Psi_{1, R} ) 
\right. \nonumber \\
&& \left. 
\hspace{2.5cm}  + i \bar{\Psi}_{2, L} \bar{\sigma}^\mu \partial_\mu \Psi_{2, L} + i \bar{\Psi}_{2, R} \sigma^\mu \partial_\mu \Psi_{2, R}
- \frac{1}{2} ( \bar{\Psi}_{2, R} \overleftrightarrow{\partial_z} \Psi_{2, L} - \bar{\Psi}_{2, L} \overleftrightarrow{\partial_z} \Psi_{2, R} ) 
\right. \nonumber \\
&& \left. 
\hspace{2.5cm} - \frac{c}{z} \left( \bar{\Psi}_{1, R} \Psi_{1, L} + \bar{\Psi}_{1, L} \Psi_{1, R}  \right) +  \frac{c}{z} \left( \bar{\Psi}_{2, R} \Psi_{2, L} + \bar{\Psi}_{2, L} \Psi_{2, R}  \right)
\right] \,.
\label{eq:5Daction}
\eeqa
The ``mass'' terms for $\Psi_1$ and $\Psi_2$ determine the scaling dimensions of the CFT operators. They are chosen to be of opposite signs because $\Psi_{1,L}$ and $\Psi_{2,R}$ should have the same scaling dimension as we see from the previous section that $\Psi_{1, L}+\Psi_{2, R}$ will correspond to the physical mesons. Their scaling dimension is $\Delta = 2-c$~\cite{Henningson:1998cd,Mueck:1998iz,Aharony:1999ti,Contino:2004vy,Cacciapaglia:2008bi}.

To incorporate the heavy quark mass we introduce the following term
\beqa
{\cal S}_{m_Q} = -M_5 \int d^5 x \left(\frac{R}{z}\right)^5 \lambda_h \, \eta \left( \bar{\Psi}_{1, L} \Psi_{2, R} + \bar{\Psi}_{2, L} \Psi_{1, R}  + h.c.\right),
\label{eq:heavymass}
\eeqa
where $\eta$ corresponds to the heavy quark scalar bilinear operator $\overline{Q}Q$. Its VEV takes the form
\beqa
\vev{\eta} = \frac{M}{\lambda_h R} \, z \,,
\eeqa
which corresponds to the heavy quark mass term (neglecting the heavy quark condensate). Plugging the VEV into Eq.~(\ref{eq:heavymass}), we obtain a \emph{constant} mass term in the AdS bulk between $\Psi_1$ and $\Psi_2$,
\beqa
{\cal S}_{m_Q} \stackrel{\vev{\eta}}{=} -M_5 \int d^5 x \left(\frac{R}{z}\right)^4  M \left( \bar{\Psi}_{1, L} \Psi_{2, R} + \bar{\Psi}_{2, L} \Psi_{1, R}  + h.c.\right).
\label{eq:heavymass2}
\eeqa
For $M > 1/R$ one might worry about the validity of the effective theory. However, this term only lifts the whole spectrum by $M$. The relevant momentum scale is still controlled by $1/L_1$ which is of order $\Lambda_{\rm QCD} < 1/R$. It is just like in the heavy-light meson system: the heavy quark simply provides a static color source and the dynamics is governed by the light quark with the relevant energy scale $\Lambda_{\rm QCD}$.

The bulk equations of motions (EOM's) are calculated to be
\beqa
&& i\bar{\sigma}^\mu \partial_\mu \Psi_{1, L} + \partial_z \Psi_{1, R} - \frac{c + 2}{z} \Psi_{1, R}  - M \Psi_{2, R} =0 \,, \\
&& i\sigma^\mu \partial_\mu \Psi_{1, R} - \partial_z \Psi_{1, L} - \frac{c - 2}{z} \Psi_{1, L}  - M \Psi_{2, L} =0 \,, \\
&& i\bar{\sigma}^\mu \partial_\mu \Psi_{2, L} + \partial_z \Psi_{2, R} + \frac{c - 2}{z} \Psi_{2, R}  - M \Psi_{1, R} =0 \,, \\
&& i\sigma^\mu \partial_\mu \Psi_{2, R} - \partial_z \Psi_{2, L} + \frac{ c + 2}{z} \Psi_{2, L}  - M \Psi_{1, L} =0 \,.
\label{eq:EOM5D}
\eeqa
If we want to calculate the spectrum and the $z$-dependent wave functions of the meson states, we need to choose boundary conditions such that the boundary terms vanish at the UV ($z=\epsilon$) and IR ($z=L_1$) boundaries. On the other hand, if we want to calculate the bulk-to-boundary propagators of the fields with which we can study the correlation functions of the HQET operators, then we need to fix $\Psi_{1,R}= \Psi_{1,R}^0$, $\Psi_{2,L}= \Psi_{2,L}^0$ at the UV boundary and introduce the following term on the UV boundary,
\beqa
{\cal L}_{\rm UV} = \frac{M_5}{2} \left( \frac{R}{\epsilon} \right)^4 \left[ \bar{\Psi}^0_{1, R} \Psi_{1, L}  - \bar{\Psi}^0_{2, L}\Psi_{2, R}  + h. c. \right]   \,,
\label{eq:UVboundary}
\eeqa
so that the total action is invariant under the variations of $\Psi_{1,L}$ and $\Psi_{2,R}$ fields.
$\bar{\Psi}^0_{1, R}$ and $\bar{\Psi}^0_{2, L}$ play the role of the sources for the operators which create the mesons. They will be discussed in more details later in subsection~\ref{sec:decayconstant-1} when we compute decay constants for the heavy-light mesons.

To solve the EOM's, it is convenient to first perform a Fourier transformation
\beq
\left(\frac{z}{R}\right)^{5/2} \psi (p, z) = \int d^4 x \, \Psi (x, z) e^{i p \cdot x}\,,
\eeq
where the additional power of $z$ is introduced for convenience of imposing boundary conditions. In the rest frame $p^\mu = (p, 0, 0, 0)$, the equations can be recombined and separated into two sets of first order differential equation in $z$. Define
\beq
\psi_a \equiv \frac{1}{\sqrt{2}} \left( \psi_{1,L} + \psi_{2,R}\right) \,, \qquad 
\psi_b \equiv \frac{1}{\sqrt{2}} \left( \psi_{2,L} - \psi_{1,R}\right) \,.
\eeq
They are coupled through their EOM's:
\beqa
\left( \partial_z - \frac{-\frac{1}{2}+c}{z}\right) \psi_b - (p-M) \psi_a = 0\,, \qquad
\left( \partial_z - \frac{-\frac{1}{2}-c}{z}\right) \psi_a + (p+M) \psi_b = 0\,.
\label{eq:first-order}
\eeqa
The first order equations can be combined to give the second order differential equation for $\psi_a$:
\beqa
\left( \partial_z - \frac{-\frac{1}{2}+c}{z}\right)\left( \partial_z - \frac{-\frac{1}{2}-c}{z}\right) \psi_a + (p^2 - M^2) \psi_a = 0 \,.
\eeqa
The second order differential equation for $\psi_b$ can be obtained by changing $c$ to $-c$. The other two combinations of fields, $\psi_c \equiv \frac{1}{\sqrt{2}} \left( \psi_{2,L} + \psi_{1,R}\right)$ and $\psi_d \equiv \frac{1}{\sqrt{2}} \left( \psi_{1,L} - \psi_{2,R}\right)$, have the same EOM's by changing $\psi_a \rightarrow \psi_c$, $\psi_b \rightarrow \psi_d$ and $c\rightarrow -c$, but are not relevant for our discussion. From Sec.~\ref{sec:match} we know that the physical meson fields map to $\psi_a$, so we will focus on the system of $\psi_a$ and $\psi_b$ only.

The solutions of $\psi_a$  and $\psi_b$ are Bessel functions:
\beqa
\psi_a(p, z) &=& c_1 \,J_\nu(kz)  + c_2 \,J_{-\nu}(kz) \, , \\
\psi_b(p, z) &=& \sqrt{\frac{{p-M}}{{p+M}}} \left[ c_1 \,J_{\nu+1}(kz)  - c_2 \,J_{-\nu-1}(kz) \right] \, ,
\eeqa
where $k\equiv \sqrt{p^2 - M^2}$ and $\nu \equiv -\frac{1}{2} - c$. If $\nu$ is an integer, the two independent solutions should be taken as $J_{|\nu|}(kz)$ and $Y_{|\nu|}(kz)$ instead. The power of the $z$ dependence of the $\Psi_a(x, z)$ in the limit of $z\rightarrow 0$ determines the scaling dimension of the corresponding heavy-light current operator.  For small $z$, we have
\beqa
z^{\frac{5}{2}} \psi_a (z) &\sim& c_1 z^{\nu + \frac{5}{2}} + c_2 z^{- \nu + \frac{5}{2}}  \sim c_1 z^{2-c} + c_2 z^{c+3}  \,, \\
z^{\frac{5}{2}} \psi_b (z) &\sim& c_1 z^{\nu + \frac{7}{2}} - c_2 z^{- \nu + \frac{3}{2}}  \sim c_1 z^{3-c} + c_2 z^{c+2}  \,.
\label{eq:scaling}
\eeqa
The scaling dimension of the operator corresponding to $\psi_a$ (sourced by $\psi_b^0$) is $\Delta = 2 - c$ for $c \leq 1/2$. For $c\geq -1/2$ there is another CFT which can be obtained by a Legendre transformation exchanging the source and the operator~\cite{Klebanov:1999tb,Contino:2004vy,Cacciapaglia:2008bi}. The operator would correspond to $\psi_b$ in that case and has the scaling dimension $c+2$, but it is not of our concern. The na\"ive dimension of the heavy-light current is 3, which would correspond to $c=-1$ and $\nu = \frac{1}{2}$. However, this current is not conserved as the corresponding symmetry is badly broken by the heavy quark mass. Therefore, there is no reason to expect $\Delta$ to remain 3. (If the heavy quark were a scalar as we have pretended it to be, the na\"ive dimension of the heavy-light current operator would be 5/2, which corresponds to $c=-\frac{1}{2}$ and $\nu=0$.) The unitarity bound requires the scaling dimension to be above the free particle limit, $\Delta >3/2$. So, in general one may expect that $3/2 < \Delta \leq 3$ which translates to $1/2 > c \geq -1$, or $-1 < \nu \leq 1/2$.

%-----------------------------------------------------------------------------
\subsection{Spectrum}
\label{sec:spectrum-1}
%-----------------------------------------------------------------------------

To obtain the spectrum of the heavy-light mesons or the corresponding 5D fermion Kaluza-Klein (KK) modes, we need to impose appropriate boundary conditions for the 5D wave functions. For the wave function to be normalizable when the UV cutoff $\epsilon$ is taken to zero, the wave function $\Psi(z)$ near $z=0$ should have a scaling power bigger than $z^{3/2}$ [or equivalently, $\psi(z)$ has a scaling power bigger than $z^{-1}$]. However, for $-1/2 < c <1/2$, this is always satisfied and the normalizability condition does not impose any extra constraint on the solutions. This is related to the fact that in this range of $c$ there are two possible CFT's discussed earlier. To pick out the CFT of our interest, we impose a stronger condition that $\psi_b(z)$ has a positive power of $z$ dependence near $z=0$, which is equivalent to the Dirichlet condition on the UV boundary for $\psi_b$. The boundary conditions for $\psi_a$ and $\psi_b$ are\footnote{If we switch the IR boundary conditions for $\psi_a$ and $\psi_b$, there would be a ``zero mode'' where $\psi_a \propto z^\nu$, $\psi_b=0$, and $p=M$. However, this is a special solution for the hard-wall model. If we imagine that the hard wall is an approximation to a soft wall, there is no solution with a soft wall which resembles that zero-mode solution.}
\beqa
\renewcommand{\arraystretch}{1.5}
\begin{array}{c|cc}
 & \mbox{UV}    &  \mbox{IR}  \\ \hline
\psi_a   &  \mbox{Mixed}   & \mbox{Dirichlet}  \\
\psi_b   &  \mbox{Dirichlet}     &  \mbox{Mixed}  \,
\end{array} 
\eeqa
where the mixed boundary condition is the generalization of the Neumann condition for the case of a warped extra dimension, which is consistent with the EOM's and the Dirichlet condition on the other component of the fermion field. The Dirichlet condition of $\psi_b$ on the UV boundary sets $c_2=0$ for $-1 \leq c < 1/2$ in the solutions. 
 The KK spectrum is then determined from the IR boundary condition:
\beqa
J_\nu (k_n L_1)  &=& 0 \,.
\eeqa
The spectrum of the KK-modes can be expressed as
\beqa
m^2_n = p_n^2 = M^2+ k_n^2 = M^2   + \frac{\left(j_{\nu,n}\right)^2}{L_1^2} \,,
\eeqa
where $j_{\nu,n}$ means the $n$'th positive zero of the Bessel function $J_\nu (x)$. In the heavy quark limit of $M > 1/L_1\sim \Lambda_{\rm QCD}$, the meson mass is linear in $M$: 
\beqa
m_n = M + {\cal O}\left( \Lambda^2_{\rm QCD}/M \right) \,.
\eeqa
Right now there is not much experimental information for higher excited modes of heavy-light mesons. As in the case of light mesons, one may not expect that the spectrum has the correct behavior for very high KK modes in this simple hard-wall model.

%-----------------------------------------------------------------------------
\subsection{Decay constants}
\label{sec:decayconstant-1}
%-----------------------------------------------------------------------------

The decay constants of meson fields can be obtained from the two-point function of the current operators which create the mesons. The two-point functions have poles corresponding to the meson masses and the decay constant of a meson is related to the residue of the corresponding pole. In AdS/QCD the two-point function can be obtained from the boundary effective action by including a source field on the UV brane and integrating out the AdS bulk using the EOM's.

Starting from the UV boundary term in Eq.~(\ref{eq:UVboundary}) and rewriting it in terms of $\psi_a$ and $\psi_b$ (ignoring $\psi_c$, $\psi_d$), we have
\beqa
{\cal L}_{\rm UV} = -\frac{M_5 \, \epsilon}{2\, R} \left(  \overline{\psi^0_b} \, \psi_a |_\epsilon + h.c. \right) \,.
\label{eq:UVboundary1}
\eeqa
As $\psi_b\sim z^{\nu+1}$ from Eq.~(\ref{eq:scaling}), in order to have a finite limit when $\epsilon$ is taken to zero, the source for the heavy-light current $h(p)$ is related to the UV boundary value $\psi_b(p, \epsilon) =\psi_b^0$ by
\beqa
h(p) = \left(\frac{R}{\alpha}\right)^{\frac{1}{2}} \left( \frac{R}{\epsilon}\right)^{-\nu -1} \psi_b (p, \epsilon)\, ,
\label{eq:source}
\eeqa
where the additional $R^{1/2}$ factor accounts for that the source $h(p)$ has engineering dimension one and $\alpha$ is expected to be an ${\cal O}(1)$ number.

To solve for $\psi_a (\epsilon)$ for the given boundary condition $\psi_b(\epsilon)=\psi_b^0$, it is convenient to define
\beqa
\zeta \equiv \begin{pmatrix} \psi_a \\ \psi_b \end{pmatrix} , \qquad \quad {\cal O} \equiv \begin{pmatrix} M & \partial_z - \frac{-\frac{1}{2}+c}{z}   \\ 
   - \partial_z + \frac{-\frac{1}{2}-c}{z} & - M 
\end{pmatrix} \,,
\eeqa
then the bulk EOM's can be written as
\beqa
{\cal O} \, \zeta =  p\, \zeta \, .
\eeqa
It is easy to show that the operator ${\cal O}$ is Hermitian with the weight function $z$ if the boundary terms vanish. The eigenfunctions of ${\cal O}$ are just the KK wave function solutions discussed in the previous subsection with eigenvalues $p_n$. If we normalize the wave functions by
\beqa
\int_{\epsilon}^{L_1} dz \, z\, \overline{\zeta}_m \zeta_n = \int_{\epsilon}^{L_1} dz\, z\, (\overline{\psi}_{a, m} \psi_{a, n} + \overline{\psi}_{b, m} \psi_{b, n}) = \delta_{mn}\, ,
\label{eq:normalization}
\eeqa
then they form an orthonormal basis which can be used to expand any function in the interval between $\epsilon$ and $L_1$. (The eigenfunctions $\psi_{a,n}$ and $\psi_{b,n}$ has dimension one in this normalization.) The solution of $\zeta (p, z)$ in the presence of the source term can be written as
\beqa
\zeta (p, z) = \sum_n \frac{c_n(p)}{p - p_n} \zeta_n (z) 
 \,,
\eeqa
and the coefficient $c_n(p)$ can be computed by
\beqa
c_n(p) &=& \int_{\epsilon}^{L_1} dz\, z\, \overline{\zeta}_n ( z) (p -p_n) \zeta( p, z) \nonumber \\
&=& z \left. (\overline{\psi}_{a, n} \psi_b - \overline{\psi}_{b, n} \psi_a) \right|_{\epsilon}^{L_1} \nonumber \\
&=& -\epsilon\, \overline{\psi}_{a, n}(\epsilon) \psi_b^0(p) \, ,
\eeqa
using the Hermiticity of ${\cal O}$.

Substituting $\zeta(p, z)$ into the boundary term in Eq.~(\ref{eq:UVboundary1}), we obtain the boundary effective action:
\beqa
{\cal L}_{\rm UV} = \frac{M_5 \epsilon^2}{ R} \,\overline{\psi^0_b} \sum_n \frac{\overline{\psi}_{a, n}(\epsilon) \psi_{a, n}( \epsilon)}{p - p_n} \, \psi^0_{b}  \,.
\eeqa
Matching the source $h(p)$ of Eq.~(\ref{eq:source}), the current-current correlator is given by
\beqa
\Pi (p) = \alpha M_5 \left( \frac{R}{\epsilon}\right)^{2\nu} \sum_n \frac{p + p_n}{p^2 - p_n^2} \left| \psi_{a, n} (\epsilon) \right|^2 \, ,
\eeqa
{}from which one can easily obtain the decay constant for the $n$th excited meson state:
\beqa
p_n^2 F_n^2 = \alpha M_5 \left( \frac{R}{\epsilon}\right)^{2\nu} 2 p_n  \left| \psi_{a, n} ( \epsilon) \right|^2 \, ,
\eeqa
or
\beqa
F_n = \sqrt{ \frac{2 \alpha M_5}{p_n} } \left( \frac{R}{\epsilon}\right)^{\nu} \left| \psi_{a, n} (\epsilon) \right| \, .
\label{eq:decayconstant}
\eeqa
The normalization condition Eq.~(\ref{eq:normalization}) for $\epsilon \to 0$ implies that\footnote{For $-1<\nu <-1/2$ ($3/2<\Delta <2$), the normalization integral is dominated by the UV region and the result will be sensitive to the UV cutoff if it is kept finite. In reality we can not expect $\epsilon/L_1 \ll 0.1$ so the applicability of the  holographic model in this range of $\nu$ may be questionable.}
\beqa
\psi_{a, n}( z) =\sqrt{\frac{p_n +M}{p_n}} \frac{J_\nu (k_n z)}{L_1 J_{\nu+1}(j_{\nu,n})} \, .
\label{eq:normalized_psi_a}
\eeqa
Expanding $\psi_{a,n}$ for small $z$, one obtains
\beqa
\psi_{a, n}(\epsilon) \approx \sqrt{\frac{p_n +M}{p_n}} \frac{1}{L_1 J_{\nu+1}(j_{\nu,n}) \Gamma(\nu +1)} \left( \frac{k_n \epsilon}{2} \right)^\nu \, .
\label{eq:psi_a_uv}
\eeqa
Substituting it into Eq.~(\ref{eq:decayconstant}), the decay constant is
\beqa
F_n = \sqrt{\frac{ 2 \alpha M_5 (p_n +M)}{p_n^2}} \frac{1}{L_1 |J_{\nu+1}(j_{\nu,n})| \Gamma(\nu +1)} \left( \frac{k_n R}{2} \right)^\nu \, .
\label{eq:decay_constant}
\eeqa
For the heavy-light meson, we have $p_n \sim M$, $k_n \sim L_1^{-1} \sim \Lambda_{\rm QCD}$. The decay constant scales as
\beqa
F_n \sim \frac{1}{\sqrt{M} }\, \frac{\sqrt{\alpha\,M_5}} {L_1}\,\,(k_n R)^\nu
\sim \frac{\sqrt{\alpha\,M_5}\, \Lambda_{\rm QCD}^{\nu +1} \, R^\nu}{\sqrt{M} }  \, .
\eeqa
The scaling $F_n \propto 1/\sqrt{M}$ agrees with the general expectation in the heavy quark limit.

%-----------------------------------------------------------------------------
\section{Parity Doubling and Chiral Symmetry Breaking}
\label{sec:parity-double}
%-----------------------------------------------------------------------------

In the chiral symmetric limit of the light flavors, the ``left-handed'' and ``right-handed'' heavy-light mesons are degenerate and form a parity doublet. After the chiral symmetry breaking effect is included, the mass eigenstates are the parity eigenstates which are linear combinations of the left-handed and right-handed fields, and the mass splitting between the parity-odd ${\cal H}$ and parity-even ${\cal H}^\prime$ states is of the order of $\Lambda_{\rm QCD}$. To discuss the parity-doublet states,
we introduce two sets of $4\times 4$ bi-spinors in the Weyl representation
\beq
{\cal H}^{\rm Weyl}_L = \begin{pmatrix} \psi_{1,L} & \psi_{2,L} \\ -\psi_{2,R} & - \psi_{1,R} \end{pmatrix},
\quad  \quad
{\cal H}^{\rm Weyl}_R = \begin{pmatrix} \phi_{1,L} & \phi_{2,L} \\ -\phi_{2,R} & - \phi_{1,R} \end{pmatrix},
\eeq
which are $(N_f, 1)$ and $(1, N_f)$ under the light flavor chiral symmetry $SU(N_f)_L\times SU(N_f)_R$. Using the relation in Eq.~(\ref{eq:HHprime}) between ${\cal H}_{L, R}$ and ${\cal H}, {\cal H}^\prime$ and rotating to the Pauli-Dirac representation, we have
\beqa
{\cal H}^{\rm PD} =\frac{1}{\sqrt{2}} \left( {\cal H}_R^{\rm PD} - {\cal H}_L^{\rm PD} \right)
= \frac{1}{2} \begin{bmatrix} -\psi_{1,L}+\psi_{2,R} + \phi_{1,L} - \phi_{2,R}
& -\psi_{2,L} +\psi_{1,R} + \phi_{2,L} - \phi_{1,R}
\\
 \psi_{1,L}+\psi_{2,R} - \phi_{1,L} - \phi_{2,R} &  \psi_{2,L}+\psi_{1,R} - \phi_{2,L} - \phi_{1,R}
 \end{bmatrix} ,
\label{eq:bi-spinor-PD-H}
\eeqa
and
\beqa
{\cal H}^{\prime \rm PD} =\frac{1}{\sqrt{2}} \left( {\cal H}_R^{\rm PD} + {\cal H}_L^{\rm PD} \right)
= \frac{1}{2} \begin{bmatrix} \psi_{1,L}-\psi_{2,R} + \phi_{1,L} - \phi_{2,R}
& \psi_{2,L} - \psi_{1,R} + \phi_{2,L} - \phi_{1,R}
\\
 -\psi_{1,L}-\psi_{2,R} - \phi_{1,L} - \phi_{2,R} &  -\psi_{2,L}-\psi_{1,R} - \phi_{2,L} - \phi_{1,R}
 \end{bmatrix}.
\label{eq:bi-spinor-PD-Hprime}
\eeqa
Similar to the simplest case in Eq.~(\ref{eq:match-dictionary}), the dictionary for relating ${\cal H}$ and ${\cal H}^\prime$ to the physical spin-0 and spin-1 states is
\beqa
H = \frac{i}{4\sqrt{M}}\, \mbox{Tr} ( -\psi_{1,L} - \psi_{2,R} + \phi_{1,L} + \phi_{2,R} ) \,,  &&H^j = \frac{1}{4\sqrt{M}} \mbox{Tr}\left[ \sigma^j (-\psi_{1,L} - \psi_{2,R} + \phi_{1,L} + \phi_{2,R}  ) \right] \,, \nonumber \\
H^\prime= \frac{i}{4\sqrt{M}}\, \mbox{Tr} ( \psi_{1,L} + \psi_{2,R} + \phi_{1,L} + \phi_{2,R} ) \,,  &&  H^{\prime j} = \frac{1}{4\sqrt{M}} \mbox{Tr}\left[ \sigma^j (\psi_{1,L} + \psi_{2,R} + \phi_{1,L} + \phi_{2,R}  ) \right] \,.
\label{eq:match-dictionary-2}
\eeqa
The parity symmetry on the $\psi$ and $\phi$ fields is defined as
\beqa
{\rm P}: \quad \vec{x} \rightarrow - \vec{x} \,, \;\sigma^\mu \leftrightarrow \bar{\sigma}^\mu \,, \;
\psi_{1,L} \leftrightarrow \phi_{2,R} \,, \; \psi_{2,R} \leftrightarrow \phi_{1,L} \,, \;
\psi_{2,L} \leftrightarrow - \phi_{1,R}\,,  \; \psi_{1,R} \leftrightarrow - \phi_{2,L} \,.
\eeqa
One can check that this is consistent with the parity of the meson fields.

Now we attempt to incorporate the parity doublet in the holographic model. The quadratic action for the ${\cal H}_R$ sector is similar to Eqs.~(\ref{eq:5Daction}) and (\ref{eq:heavymass2}):
\beqa
{\cal S}_{5D} &\supset& M_5 \int d^5 x \left(\frac{R}{z} \right)^4 \left[
i \bar{\Phi}_{1, L} \bar{\sigma}^\mu \partial_\mu \Phi_{1, L} + i \bar{\Phi}_{1, R} \sigma^\mu \partial_\mu \Phi_{1, R}
- \frac{1}{2} ( \bar{\Phi}_{1, R} \overleftrightarrow{\partial_z} \Phi_{1, L} - \bar{\Phi}_{1, L} \overleftrightarrow{\partial_z} \Phi_{1, R} ) 
\right. \nonumber \\
&& \left. 
\hspace{2.5cm}  + i \bar{\Phi}_{2, L} \bar{\sigma}^\mu \partial_\mu \Phi_{2, L} + i \bar{\Phi}_{2, R} \sigma^\mu \partial_\mu \Phi_{2, R}
- \frac{1}{2} ( \bar{\Phi}_{2, R} \overleftrightarrow{\partial_z} \Phi_{2, L} - \bar{\Phi}_{2, L} \overleftrightarrow{\partial_z} \Phi_{2, R} ) 
\right. \nonumber \\
&& \left. 
\hspace{-1.3cm} - \frac{c}{z} \left( \bar{\Phi}_{1, R} \Phi_{1, L} + \bar{\Phi}_{1, L} \Phi_{1, R}  \right) +  \frac{c}{z} \left( \bar{\Phi}_{2, R} \Phi_{2, L} + \bar{\Phi}_{2, L} \Phi_{2, R}  \right)
- M \left( \bar{\Phi}_{1, L} \Phi_{2, R} + \bar{\Phi}_{2, L} \Phi_{1, R}  + h.c.\right)  \right] \,.
\label{eq:5Daction2}
\eeqa
The parameters $M$ and $c$ for the $\Phi$ fermions take the same values as the action in Eqs.~(\ref{eq:5Daction}) and (\ref{eq:heavymass2}) to preserve the parity symmetry of the total action:
\beqa
{\rm P}: \quad \vec{x} \rightarrow - \vec{x} \,, \;\sigma^\mu \leftrightarrow \bar{\sigma}^\mu \,, \;
\Psi_{1,L} \leftrightarrow \Phi_{2,R} \,, \; \Psi_{2,R} \leftrightarrow \Phi_{1,L} \,, \;
\Psi_{2,L} \leftrightarrow - \Phi_{1,R}\,,  \; \Psi_{1,R} \leftrightarrow - \Phi_{2,L} \,.
\label{eq:parity-double}
\eeqa

The chiral symmetry breaking in AdS/QCD is parametrized by the VEV of a bi-fundamental scalar field, $\Sigma$ $(N_f, \bar{N}_f)$, under $SU(N_f)_L\times SU(N_f)_R$, 
\beqa
\langle \Sigma (z) \rangle  = \frac{M_q}{R}\,z + \frac{\xi}{R L_1^3}\,z^3 \,,
\eeqa
where we followed the notation in Ref.~\cite{DaRold:2005zs} and assumed the na\"ive scaling dimension for the corresponding $q\bar{q}$ operator.
$M_q$ can be matched to the bare quark mass in the chiral Lagrangian but will be neglected in the rest of this paper. Under the parity transformation, we have $\Sigma \rightarrow \Sigma^\dagger$. The chiral symmetry breaking in the holographic model of heavy-light mesons can be induced by the coupling of $\Psi$ and $\Phi$ fields to $\Sigma$:
\beqa
S_{\rm int.} &\supset&  M_5 \int d^5 x \left(\frac{R}{z} \right)^5 \left[ - \lambda\,\left( \bar{\Psi}_1 \Sigma \Phi_2 + \bar{\Psi}_2 \Sigma \Phi_1 \,+\, h.c.   \right) \right]   \nonumber \\
&=& M_5 \int d^5 x \left(\frac{R}{z} \right)^5 \left[ - \lambda\,\left( \bar{\Psi}_{1, L} \Sigma \Phi_{2, R} +  \bar{\Psi}_{1, R} \Sigma \Phi_{2, L} + \bar{\Psi}_{2, L} \Sigma \Phi_{1, R} + \bar{\Psi}_{2, R} \Sigma \Phi_{1, L}  \,+\, h.c.   \right) \right]  \,, 
\label{eq:interaction}
\eeqa
where $\lambda$ can be chosen real by a phase rotation to be consistent with the parity symmetry.
The action is invariant under the $SU(N_f)_L\times SU(N_f)_R$ flavor symmetry and also the parity transformation of Eq.~(\ref{eq:parity-double}). Substituting in  the VEV of the $\Sigma$ field and defining $\sigma(z) = \lambda \,\xi\, z^2/L^3_1$, we obtain
\beqa
S_{\rm int.} &\supset&  - M_5 \int d^5 x \left(\frac{R}{z} \right)^4 
\sigma(z)   \left( \bar{\Psi}_{1, L}  \Phi_{2, R} +  \bar{\Psi}_{1, R}  \Phi_{2, L} + \bar{\Psi}_{2, L} \Phi_{1, R} + \bar{\Psi}_{2, R}  \Phi_{1, L}  \,+\, h.c.   \right) \nonumber  \\
&=&  M_5 \int d^5 x \left(\frac{R}{z} \right)^4 
\sigma(z)\,  \Tr \left( \overline{\cal H}^{\rm PD}_L {\cal H}^{\rm PD}_R \,+\, h.c.   \right) \, .
\eeqa
The $\sigma (z)$ serves as an off-diagonal mass term which split the parity-even and parity-odd states.

The EOM's can be similarly obtained by performing the Fourier transformations on both $\Psi$ and $\Phi$ fields,
\beqa
\left(\frac{z}{R}\right)^{5/2} \psi_{1, 2} (p, z) &=& \int d^4 x \, \Psi_{1, 2}(x, z) e^{i p \cdot x}\,, \\
\left(\frac{z}{R}\right)^{5/2} \phi_{1, 2} (p, z) &=& \int d^4 x \, \Phi_{1, 2}(x, z) e^{i p \cdot x}\,. 
\eeqa
As in Sec.~\ref{sec:5D-lagrangian}, we first define
\beqa
&&\psi_a = \frac{1}{\sqrt{2}} ( \psi_{1,L} + \psi_{2, R} ) \,, \quad \psi_b = \frac{1}{\sqrt{2}} ( \psi_{2,L} - \psi_{1, R} ) \,,  \nonumber \\
&&\phi_a = \frac{1}{\sqrt{2}} ( \phi_{1,L} + \phi_{2, R} ) \,, \quad \phi_b = \frac{1}{\sqrt{2}} ( \phi_{2,L} - \phi_{1, R} )  \,,
\eeqa
then the parity-odd and parity-even states can be identified as
\beqa
{\mbox{P-odd}}: &&\qquad  u_a = \frac{1}{\sqrt{2}} ( \psi_a - \phi_a) \,,\qquad u_b = \frac{1}{\sqrt{2}} ( \psi_b - \phi_b)  \,, \\
{\mbox{P-even}}: &&\qquad  u^\prime_a = \frac{1}{\sqrt{2}} ( \psi_a + \phi_a) \,,\qquad u^\prime_b = \frac{1}{\sqrt{2}} ( \psi_b + \phi_b)   \,.
\eeqa
The EOM's for P-odd and P-even states are given by
\beqa
&&\left( \partial_z - \frac{-\frac{1}{2}+c}{z}\right) u_b - \left[p-M + \sigma(z)\right] u_a = 0\,, \;
\left( \partial_z - \frac{-\frac{1}{2}-c}{z}\right) u_a + \left[p+M - \sigma(z)\right] u_b = 0\,, 
\label{eq:double-odd-EOM} \\
&&\left( \partial_z - \frac{-\frac{1}{2}+c}{z}\right) u^\prime_b - \left[p-M - \sigma(z)\right] u^\prime_a = 0\,, \;
\left( \partial_z - \frac{-\frac{1}{2}-c}{z}\right) u^\prime_a + \left[p+M + \sigma(z)\right] u^\prime_b = 0\,.
\label{eq:double-even-EOM}
\eeqa

The spectrum and decay constants can be solved by imposing appropriate boundary conditions in the same ways as in Sec.~\ref{sec:5D-lagrangian}. However, if $\lambda$ is a constant, $\sigma (z)$ grows as $z^2$ in the IR. It renders an attractive potential in the Schr\"{o}dinger-like equation for the parity-odd states and a repulsive potential for the parity-even states in the IR. As a result, the wave functions for the low-lying parity-odd states grow exponentially in the IR region while the parity-even states are repelled away from the IR. After normalizing the wave functions, the wave functions of the parity-odd states become highly suppressed on the UV boundary, which implies that the decay constants of the parity-odd states are also highly suppressed (relative to those of parity-even states). This is in conflict with results from lattice simulations and other calculations. We find that a constant $\sigma (z)$ provides a better description of the experimental and lattice data. This would require that the coupling $\lambda$ scales as $1/z^2$. Such a $z$-dependence corresponds to an explicit conformal breaking dimension-six operator. That is, one can imaging that $\lambda$ itself arises from a VEV of a scalar field which corresponds to the dimension-six current-current interaction term. It is reminiscent of Ref.~\cite{Bardeen:1993ae} which approximates the QCD by a Nambu-Jona-Lasinio model~\cite{Nambu:1961tp,Nambu:1961fr} with a local current-current interaction replacing the effects of gluon exchanges. Usually these high-dimensional operators only modify physics in the UV and their effects on the bulk solutions diminish as negative powers of $z$. However, here the $1/z^2$ dependence is multiplied by the growing $\Sigma$ VEV, so it's modification on the bulk EOM's is constant instead of vanishing towards the IR. We do not have a good justification why the constant $\lambda$ term, which is allowed by symmetry, is absent or suppressed relative to the $1/z^2$ term which would come from a higher dimensional operator in the 5D theory. We will simply assume that $\lambda \propto 1/z^2$ guided by the experimental or lattice data in our following analysis.

%-----------------------------------------------------------------------------
\subsection{Mass Spectrum and Decay Constants for a Constant $\sigma(z)$}
\label{sec:constantsigma}
%-----------------------------------------------------------------------------
We assume that the coupling $\lambda$ takes the following form:
\beqa
\lambda= \bar{\lambda} \frac{R^2}{z^2},
\eeqa
where $\bar{\lambda}$ is a constant. The chiral symmetry breaking mass $\sigma(z)$ becomes a constant:
\beqa
\sigma (z) = \frac{\bar{\lambda} \xi R^2}{L_1^3} \equiv \sigma \, .
\eeqa
By examining the Eqs.~(\ref{eq:double-odd-EOM}) and (\ref{eq:double-even-EOM}) one can see that one only needs to replace $M$ in the results of Sec.~\ref{sec:5D-lagrangian} by $M-\sigma$ ($M+\sigma$) for the parity-odd (-even) states. The solutions are the same Bessel functions and the mass spectra for the parity-odd and even states are given in terms of roots of Bessel functions,
\beqa
m^{-}_n = p^{\rm odd}_n &=& \sqrt{ (M - \sigma)^2+ k_n^2}= \sqrt{ (M -\sigma)^2   + \frac{\left(j_{\nu,n}\right)^2}{L_1^2}  } \,, \\
m^{+}_n = p^{\rm even}_n &=& \sqrt{ (M + \sigma)^2+ k_n^2}= \sqrt{ (M + \sigma)^2   + \frac{\left(j_{\nu,n}\right)^2}{L_1^2}  } \,. 
\eeqa
For $\sigma \sim 1/L_1 \ll M$, the inter-multiplet mass splitting is 
\beqa
\Delta M = m^{\rm even}_n - m^{\rm odd}_n \approx \sigma \approx \Lambda_{\rm QCD} \,,
\eeqa
which is the same as in Ref.~\cite{Bardeen:2003kt}. 

The decay constants are similarly obtained by replacing $M$ by $M\mp \sigma$ from Eq.~(\ref{eq:decay_constant}),
\beqa
F_n^{\mp} = \sqrt{\frac{ 2 \alpha M_5 (p_n +M\mp \sigma)}{p_n^2}} \frac{1}{L_1 |J_{\nu+1}(j_{\nu,n})| \Gamma(\nu +1)} \left( \frac{k_n R}{2} \right)^\nu \, , 
\label{eq:decay_constant_double}
\eeqa
where $- (+)$ denotes parity-odd (-even) states.

To compare with the experimental and lattice data, we use $L_1^{-1} = m_\rho/ j_{0,1} = 322$~MeV from fitting the $\rho$ mass and $M_5 R = N_c/ (12 \pi^2)$ from matching to perturbative QCD~\cite{DaRold:2005zs}, which provide a good fit to the light meson data. Because the hyperfine splitting between the spin-0 and spin-1 states is ${\cal O}(1/M)$ effect which is not included in our model, for the $D$ mesons we take the weighted average of the measured masses of the spin-0 and spin-1 mesons, $(3 m^{J=1}+ m^{J=0})/4$, as the experimental value to remove the hyperfine effect. For the $B$ mesons, only the $1^+$ state is identified experimentally for the lowest parity-even states, so we only fit spin-1 meson masses. The hyperfine splittings in $B$ mesons are smaller anyway. For the decay constants, it is known that the ratio between the $D$ and $B$ mesons, $F_D/F_B$, does not agree well with the $\sqrt{M_B/M_D}$ scaling relation. The reason could be attributed to that the charm quark is not heavy enough (compared to $\Lambda_{\rm QCD}$) to be in the heavy quark limit. Given that, we list two different fits for the decay constants of the $D$ mesons: one with the same combination of $\alpha$ and $R$ values derived from the $B$ meson decay constants and the other with parameters fit directly to the $D$ meson data. As a comparison to the experimental and lattice data, we show the numerical results of the spectra and decay constants of the $B$ and $D$ mesons in Tables~\ref{tab:mesonConstantFit1} and \ref{tab:mesonConstantFit2} for $\nu=0$ and $\nu=-1/2$ as specific examples.
\begin{table}[t!]
\vspace*{4mm}
\renewcommand{\arraystretch}{1.5}
\centerline{
\begin{tabular}{|c|c|cc|cc|cc|}
\hline \hline
 & & \multicolumn{2}{c|}{Exp.\ or lattice values} & \multicolumn{2}{c|}{Fit to $B$} & \multicolumn{2}{c|}{Fit to $D$} \\ \hline
$\mbox{State}$ & $J^P$  &   Mass & $F$ & Mass & $F$   & Mass & $F$ \\ \hline \hline \hline
$B^*$   &  $1^-$    &    5325   & 186(4)  & [5325] &  [186]  & $-$ & $-$ \\ \hline
$B_1$   &  $1^+$   &  5724(2)    & ?  & [5724] &  179  & $-$ & $-$ \\ \hline \hline
$?$   &  $1^-$  &    ?  & ? &   5560   & 275  & $-$ & $-$ \\ \hline
$?$   &  $1^+$  &  ?    & ? &  5943    &  266  & $-$ & $-$ \\ \hline \hline \hline
$D^0, D^*$   &  $0^-, 1^-$   & $1971$  & 209(5)  &    [1971]  & 300 &  [1971] & [209]  \\ \hline
$D^*_0, D_1^0$   &  $0^+, 1^+$   & $2400(34)$  & 200(50)  &    2347  & 277 & [2400] & 191 \\ \hline \hline
$D(2550)^0$   &  $0^-$   & $2539(8)$  & ?  &    2539  & 381 & 2539 & 265  \\ \hline
$?$   &  $0^+$   & $?$  &   ? &  2840  & 368 & 2884 & 254  \\ \hline \hline
\end{tabular}
}
\caption{The numerical values of the masses and decay constants of the lowest and first excited states of the $0^-(1^-)$ and $0^+(1^+)$ mesons from AdS-HQET for $\nu=0\,(\Delta=5/2)$. All numbers are in MeV. The experimental mass values are taken from the Particle Data Group~\cite{Beringer:1900zz}. For $D$ mesons we use the weighted average of the spin-zero and spin-one meson masses. There are two mass-close $1^+$ states, $D_1 (2420)^0$ and $D_1 (2430)^0$. We choose $D_1 (2430)^0$ because it has the similar width to that of $D^*_0$. The numbers inside square brackets are input values to fit for the model parameters. Because $B$ mesons are expected to be better described by HQET, we fit $B$ meson data first to determine the model parameters. The resulting model parameters are $M=5470$~MeV for $B$ mesons, $M=2014$~MeV for $D$ mesons, and $\sigma = 201$~MeV.  The decay constants from lattice calculations are used to fix other input parameter values. We choose the latest results from the HPQCD Collaboration ($F_B$)~\cite{Dowdall:2013tga}, the MILC Collaboration ($F_D$)~\cite{Bazavov:2012dg}, and the UKQCD Collaboration ($F_{D_{0^+}}$)~\cite{Herdoiza:2006qv}. The fit value for $\alpha/R$ from $F_B$ is 4738~MeV. We also perform a fit to the $D$ meson data only, which is listed in the last two columns. The corresponding model parameters are $M=2042$~MeV, $\sigma= 230$~MeV and $\alpha/R= 2301$~MeV. 
\label{tab:mesonConstantFit1}} 
\end{table}
\begin{table}[t!]
\vspace*{4mm}
\renewcommand{\arraystretch}{1.5}
\centerline{
\begin{tabular}{|c|c|cc|cc|cc|}
\hline \hline
 & & \multicolumn{2}{c|}{Exp.\ or lattice values} & \multicolumn{2}{c|}{Fit to $B$} & \multicolumn{2}{c|}{Fit to $D$} \\ \hline
$\mbox{State}$ & $J^P$  &   Mass & $F$ & Mass & $F$   & Mass & $F$ \\ \hline \hline \hline
$B^*$   &  $1^-$    &    5325   & 186(4)  & [5325] &  [186]  & $-$ & $-$ \\ \hline
$B_1$   &  $1^+$   &  5724(2)    & ?   & [5724] &  179  & $-$ & $-$ \\ \hline \hline
$?$   &  $1^-$       &    ?  & ? &   5514   & 181  & $-$ & $-$ \\ \hline
$?$   &  $1^+$      &  ?    & ? &  5900    &  175  & $-$ & $-$ \\ \hline \hline \hline
$D^0, D^*$   &  $0^-, 1^-$   & $1971$  & 209(5)  &    [1971]  & 303 &  [1971] & [209]  \\ \hline
$D^*_0, D_1^0$   &  $0^+, 1^+$   & $2400(34)$  & 200(50)  &    2361  & 278 & [2400] & 190 \\ \hline \hline
$D(2550)^0$   &  $0^-$   & $2539(8)$  & ?  &    2435  & 260 & 2435 & 179  \\ \hline
$?$   &  $0^+$   & $?$  &   ? &  2760  & 248 & 2794 & 170  \\ \hline \hline
\end{tabular}
}
\caption{Same as Table~\ref{tab:mesonConstantFit1} but for $\nu=-1/2\,(\Delta=2)$.  The fit model parameters are $M=5501$~MeV for $B$ mesons, $M=2105$~MeV for $D$ mesons, and $\sigma = 200$~MeV.  The fit value for $\sqrt{\alpha}/R$ from $F_B$ is 2379~MeV. For the fit to $D$ mesons only, the corresponding model parameters are $M=2126$~MeV, $\sigma= 221$~MeV and $\sqrt{\alpha}/R= 1638$~MeV. 
\label{tab:mesonConstantFit2}} 
\end{table}

In the hard-wall holographic model, the mass splittings among KK modes of the same parity scale as $\Lambda_{\rm QCD}^2/M$. Experimentally, not many excited heavy-light mesons have been identified with clear quantum numbers. In the particle listing of Particle Data Group~\cite{Beringer:1900zz}, there is a $D(2550)^0$ with $J^P=0^-$ which may be identified as the excited state of $D^0$. We see from Table~\ref{tab:mesonConstantFit1} that $\nu=0$ gives a good fit to its mass while $\nu=-1/2$ prediction is low. For the decay constants, there are many good determinations of those of the lowest odd parity states.  In addition to the latest lattice results listed in the Tables, the CLEO experiment has measured the branching fraction of $D^+ \to \mu^+ \nu$ which translates to a value $f_D= 206.7\pm 8.5 \pm 2.5$~MeV~\cite{Eisenstein:2008aa}. QCD sum rules also provide similar values~\cite{Colangelo:1991ug,Lucha:2011zp}. The phenomenological calculations of the decay constants of the lowest parity-even states, on the other hand, are scattered in a wide range~\cite{Colangelo:1991ug,Colangelo:1992kc,Veseli:1996yg,Morenas:1997rx,Zhu:1998wy,Cheng:2003sm,Jugeau:2005yr}. These calculations are also divided into determinations at the finite charm mass and in the heavy quark limit, because the charm quark is not very heavy compared to $\Lambda_{\rm QCD}$.  A summary of calculated results from different approaches can be found in Ref.~\cite{Jugeau:2005yr}. For the decay constant in the heavy quark limit, the more recent determination from an unquenched lattice QCD calculation of the UKQCD Collaboration~\cite{Herdoiza:2006qv} has $f_P^{\rm static}= 294(88)$~MeV. It may be compared with our $f_D$'s obtained from fitting the $B$ meson data which is closer to that limit. Less is known about the decay constants of the excited heavy-light meson states. There is a lattice calculation of the decay constants of $B_s$ and its first radially excited state $B'_s$ using the quenched approximation~\cite{Blossier:2010mk}. The result is
\beqa
\frac{f_{B'_s}^{\rm stat} \sqrt{m_{B'_s}}}{f_{B_s}^{\rm stat} \sqrt{m_{B_s}}}= 1.24(7)\,,
\eeqa
which is not far from our results for $B_1$ and $B^*$. 

\begin{figure}[ht!]
\begin{center}
\includegraphics[width=0.6\textwidth]{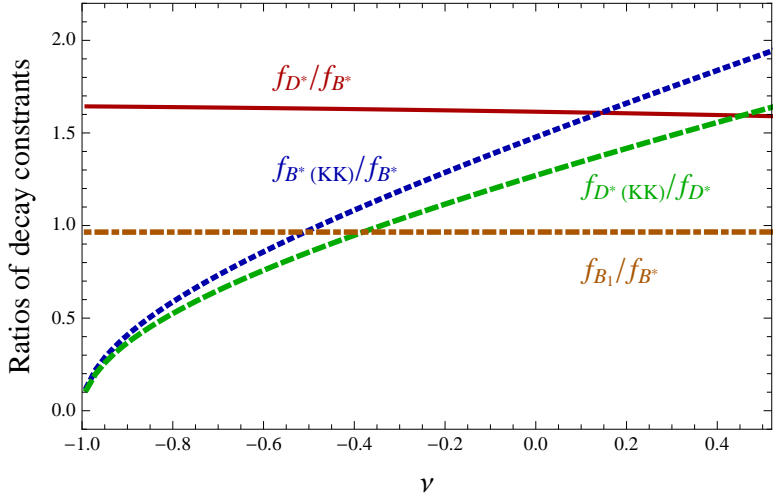} 
\caption{Ratios of decay constants as a function of $\nu = \Delta - 5/2$. For each $\nu$, we choose the model parameters to fit $m_{B^*}$, $m_{B_1}$, $m_{D^0, D^*}$ and $f_{B^*}$.
}
\label{fig:decay_constants}
\end{center}
\end{figure}

In Fig.~\ref{fig:decay_constants}, we display ratios of heavy-light meson decay constants obtained in the holographic model as a function of $\nu$. Only the ratios between the KK excited states and the ground states have a significant dependence on $\nu$. More data on the excited states may help to pin down the preferred range of $\nu$. Currently $\nu\approx 0$ seems to provide a reasonable fit.
%

%-----------------------------------------------------------------------------
\subsection{Couplings between Heavy-light Mesons and Pions}
\label{sec:couplings}
%-----------------------------------------------------------------------------

In HQET, the coupling between the heavy-light mesons of opposite parities and the pion is related to the mass splitting between the the parity-even and parity-odd mesons and the pion decay constant by the Goldberger-Treiman relation~\cite{Bardeen:2003kt}:
\beqa
\Delta M = g_\pi F_\pi \, ,
\label{eq:goldberger-treiman}
\eeqa
at the leading order, where $F_\pi = 92.2$~MeV is the pion decay constant. The coupling can receive corrections and in Ref.~\cite{Bardeen:2003kt} the corrections are parametrized by a phenomenological parameter $G_A$ which is expected to be ${\cal O}(1)$. The partial widths of the decay $0^+ (1^+) \rightarrow 0^- (1^-) + \pi$ can then be calculated in terms of $G_A$.

In the holographic model, the coupling can be calculated from the overlap integral of the wave functions of the fermions and the pion along the $z$ direction, which in the dual picture includes the leading $N_c$ corrections. In AdS/QCD, the pion field is contained in both the scalar field $\Sigma$ and the fifth component of the axial vector field $A_5$. In terms of the Goldstone bosons, the scalar field can be expressed as 
\beqa
\Sigma = \langle \Sigma \rangle\, e^{i\,P/  \langle \Sigma \rangle} \equiv  v(z)\, e^{i\,P/ v(z) }  \,,
\eeqa
where $P= P_A T_A$ contains the associated broken generators $T_A$. [In our normalization, the generators are normalized to $\Tr (T_A T_B) = \delta_{AB}/2$.]
The vector and axial-vector gauge fields of the light flavor symmetry are related to the left-handed and right-handed gauge fields by
\beqa
V_M &=& \frac{1}{\sqrt{2} } (R_M + L_M) \,,\nonumber \\
A_M &=& \frac{1}{\sqrt{2} } (R_M - L_M) \,.
\label{eq:gauge-field-relation}
\eeqa
In the unitary gauge, the uneaten Nambu-Goldstone bosons are~\cite{DaRold:2005zs}
\beqa
P = - \frac{z^3}{\sqrt{2} \,R^2 \,v(z)} \partial_5 \left(\frac{A_5}{z}\right) = - \frac{L_1^3}{\sqrt{2} \,R \,\xi} \partial_5 \left(\frac{A_5}{z}\right)\,. 
\eeqa

The interactions of the Nambu-Goldstone bosons and the bulk fermion fields can come from the Yukawa interaction terms in Eq.~(\ref{eq:interaction}) and the gauge interactions from promoting the derivatives in the kinetic terms in Eqs.~(\ref{eq:5Daction}) and (\ref{eq:5Daction2}) to covariant derivatives. Expanding the Yukawa interaction
\beqa
S_{\rm int.} &=& M_5 \int d^5 x \left(\frac{R}{z} \right)^5 \left[ - \lambda\,\left( \bar{\Psi}_{1, L} \Sigma \Phi_{2, R} +  \bar{\Psi}_{1, R} \Sigma \Phi_{2, L} + \bar{\Psi}_{2, L} \Sigma \Phi_{1, R} + \bar{\Psi}_{2, R} \Sigma \Phi_{1, L}  \,+\, h.c.   \right) \right]  \,. 
\eeqa
to linear order in $P$, we obtain the couplings
\beqa
S_{\rm Yuk.} &=& M_5 \int d^5 x \left(\frac{R}{z} \right)^5 \, (- i \,\lambda) \, 
\left[ \overline{U}_a  \,P \, U^\prime_a - \overline{U}_b  \,P \, U^\prime_b \right] \,+\, h.c. \,,
\label{eq:pion_coupling}
\eeqa
where $U_{a, b}$ ($U^\prime_{a, b}$) are the five-dimensional field associated with the parity-odd (-even) states, i.e.,
\beqa
U_a(x, z)= \frac{1}{\sqrt{2}} \left[ \Psi_a (x, z) - \Phi_a (x,z)\right] = \frac{1}{2} \left[ \Psi_{1,L}(x,z)+\Psi_{2,R}(x,z)- \Phi_{1,L}(x,z)-\Phi_{2,R}(x,z)\right] \, ,
\eeqa
and so on.

From the gauge interactions and keeping only the $L_5$ or $R_5$ parts, we have 
\beqa
S_{\rm gauge} &=& M_5 \int d^5 x \left(\frac{R}{z} \right)^4 \, (- i) \, 
\left( \overline{\Psi}_{1, R}\,L_5\,\Psi_{1, L} - \overline{\Psi}_{1, L}\,L_5\,\Psi_{1, R} +  \overline{\Psi}_{2, R}\,L_5\,\Psi_{2, L} - \overline{\Psi}_{2, L}\,L_5\,\Psi_{2, R} \right. \nonumber \\
&& \left.  \qquad\qquad \quad +\, \overline{\Phi}_{1, L}\,R_5\,\Phi_{1, R} - \overline{\Phi}_{1, R}\,R_5\,\Phi_{1, L} +  \overline{\Phi}_{2, L}\,R_5\,\Phi_{2, R} - \overline{\Phi}_{2, R}\,R_5\,\Phi_{2, L} 
 \right) \, \nonumber \\
&\supset&
M_5 \int d^5 x \left(\frac{R}{z} \right)^4 \, i \frac{1}{\sqrt{2}} 
\left[ \overline{U}_b\,A_5\,U_a^\prime - \overline{U}_a\,A_5\,U_b^\prime  \right] \,+\, h.c. \,.
\eeqa
As we discussed previously the heavy-light mesons correspond to the $U_a$ and $U'_a$ fields, while $U_b$ and $U'_b$ can at most contribute through mixings which are suppressed by $1/M$, the dominant contribution comes from the first term in Eq.~({\ref{eq:pion_coupling}). We can separate the 5D fields into products of the 4D fields and the corresponding wave functions in the fifth dimension,
\beqa
U_a(x, z) &=& {\cal H}_n (x)\left(\frac{z}{R}\right)^{5/2} f_{a,n}(z)\,, \\
A_5 (x, z) &=& \boldsymbol \pi(x)\, f_0^\pi (z) = \pi_A(x)\,T_A\,f_0^\pi (z) \,  ,
\eeqa
where
$
f_{a, n}(z) \propto u_a(p_n, z) \, ,
$
and from Ref.~\cite{DaRold:2005zs}
\beqa
f_0^\pi (z) &\propto& \frac{z^3}{L_1^3} \left[ I_{2/3} \left( \frac{\sqrt{2} \xi}{3} \frac{z^3}{L_1^3}\right) - \frac{ I_{2/3}(\sqrt{2} \xi/3)}{K_{2/3}(\sqrt{2}\xi/3)}  K_{2/3} \left( \frac{\sqrt{2} \xi}{3} \frac{z^3}{L_1^3}\right) \right]\, .
\eeqa
The normalizations are chosen such that the 4D fields are canonically normalized,
\beqa
&& M_5 \int_\epsilon^{L_1} dz \left(\frac{z}{R}\right) f_{a, n}^2 =1 \, , \quad \mbox{and }  \left. 
-\frac{M_5\,z^2}{2\,R\,v(z)^2} f_0^\pi \partial_z \left( \frac{1}{z}\,f_0^\pi\right) \right|_\epsilon=1 \, .
\eeqa
If we take the 5D coupling to be proportional to $1/z^2$ as discussed earlier in this section,
\beqa
\lambda = \frac{\sigma L_1^3}{\xi z^2}\,,
\eeqa
then the wave function of the lowest parity-odd and -even states are the same:
\beqa
f_{a,1}(z) = f'_{a,1}(z) =\frac{\sqrt{2}}{\sqrt{M_5 R}} \frac{R}{L_1} \frac{ J_\nu ( j_{\nu ,1} z/L_1)}{J_{\nu+1} ( j_{\nu, 1})} \, .
\eeqa
The coupling of the interaction between the parity-even and parity-odd heavy-light mesons and the pion, $\left[ - i g_\pi \Tr (\overline{\cal H} \boldsymbol \pi {\cal H}^\prime) + h.c.\right]$, 
 is given by\footnote{The factor $1/4$ in Eq.(12)  of Ref.~\cite{Bardeen:2003kt} is due to the $\sqrt{2}$ different normalization of the heavy-light bi-spinor field and a factor of 2 in the definition of $\Sigma$.}
\beqa
g_\pi &=& M_5 \int_\epsilon^{L_1} dz\,  (-\lambda) f_{a,1}^2 \left[ -\frac{L_1^3}{\sqrt{2} \,R\, \xi} \partial_z \left(\frac{1}{z} f_0^\pi\right)\right]  \nonumber \\
&=& \frac{\sigma L_1}{\sqrt{M_5 R}} \times \mbox{ an ${\cal O}(1)$ numerical factor} \, .
\label{eq:intermultiplet_coupling}
\eeqa 
The couplings for different choices of $\nu$ are shown in Fig.~\ref{fig:coupling}.
\begin{figure}[ht!]
\begin{center}
\includegraphics[width=0.6\textwidth]{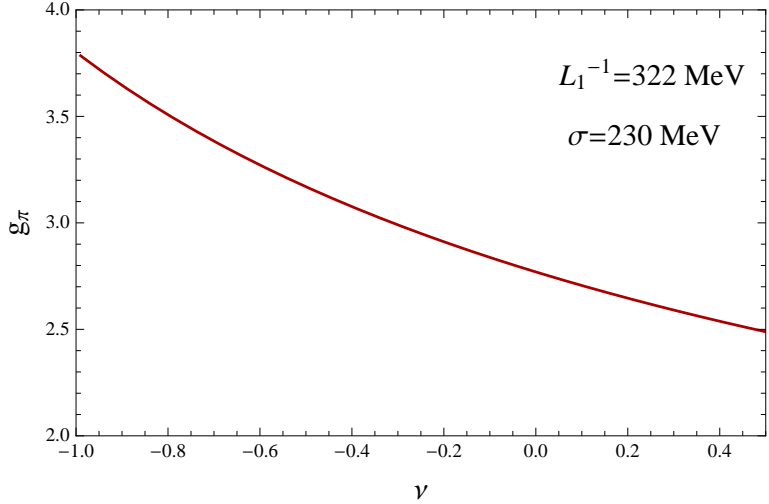} 
\caption{The dependence of the coupling $g_\pi$ on $\nu = \Delta-5/2$, where we take $L_1^{-1}=322$~MeV, $\sigma = 230$~MeV, $\sqrt{M_5 R}= 1/(2\pi)$. For $\nu <-1/2$, the integral Eq.~(\ref{eq:intermultiplet_coupling}) is dominated by the UV region so the result will be sensitive to the UV cutoff $\epsilon$ if it is kept finite.
}
\label{fig:coupling}
\end{center}
\end{figure}
In comparison, the lowest order Goldberger-Treiman relation $g_\pi = \Delta M/ F_\pi$ would give $g_\pi= 4.33\, (4.65)$ if one uses the $B\, (D)$ meson spectrum given in Table~\ref{tab:mesonConstantFit1}. We see that in the AdS-HQET model the coupling is smaller than that given by the Goldberger-Treiman relation, which implies that the leading $N_c$ corrections reduce the coupling. The coupling was also calculated using QCD sum rules~\cite{Colangelo:1995ph,Colangelo:1997rp,Dai:1998ve,Zhu:1998wy}. In Ref.~\cite{Zhu:1998wy} the coupling of $B'_0 B \pi$ was calculated to be $2.8\pm 0.5$ by including the $B\pi$ continuum contribution. It is close to our value $g_\pi =2.77$ at $\nu=0$.

With the coupling we can calculate the decay rate of the inter-multiplet transition $(0^+,\, 1^+) \to (0^-, \, 1^-) + \pi$. To do that let us consider the simplest $0^+ 0^- \pi^0$ tri-scalar coupling. From Eq.~(\ref{eq:match-dictionary-2}) the (dimension-1) spin-0 meson field is embedded in bi-spinor ${\cal H}$ field as
\beqa
{\cal H^{\rm PD}} \supset \begin{pmatrix}  - & - \\ i \sqrt{M} \,H \, \mathbb{I}_2 & - \end{pmatrix} \, .
\eeqa
The neutral Goldstone boson $\pi^0$ is accompanied by the generator $\sigma_3 /2$. One can see that the coupling among $0^+ 0^- \pi^0$ is $i g_\pi M$ after performing the trace. The total decay width of the $D_0^{*}$ to $D\,\pi$ is 3 times the decay width to $D^0\, \pi^0$:
\beqa
\Gamma (D_0^*)=  3\times \frac{1}{8 \pi} | {\cal M}|^2 \frac{|\vec{p}_\pi|}{M_{D_0^*}^2} = \frac{3 g_\pi^2}{8 \pi} \frac{M^2}{M_{D_0^*}^2} | \vec{p}_\pi|\, ,
\eeqa
where 
\beqa
|\vec{p}_\pi| = \frac{[ (M_{D^*}^2 - (M_D+m_\pi)^2) (M_{D^*}^2 - (M_D-m_\pi)^2)]^{1/2}}{2 M_{D^*}} \, .
\eeqa
Using $M_{D_0^*} = 2318$~MeV, $M_D= 1865$~MeV, $m_\pi=135$~MeV, we obtain $|\vec{p}_\pi|= 390$~MeV. Taking $M\approx(M_{D_0^*}+M_D)/2$, we have
\beqa
\Gamma(D_0^*) = 37.9\, g_\pi^2 \mbox{ MeV} = 291 \left(\frac{g_\pi}{2.77}\right)^2 \mbox{ MeV} \, ,
\eeqa
which agrees well with the experimental value $267\pm40$~MeV from the Particle Data Group~\cite{Beringer:1900zz}, assuming that the total width is dominated by the single pion decay.

The intra-multiplet coupling between the $0^-$, $1^-$, and the pion can be written down by combining HQET with the chiral perturbation theory~\cite{Wise:1992hn,Burdman:1992gh,Yan:1992gz}. In our model, it could arise from the magnetic dipole moment operator in 5D:
\beqa
{\cal S}_{\rm dipole} &=& M_5 R \int d^5 x \left( \frac{R}{z}\right)^3 \lambda_D \bigg[ \overline{\Psi}_1 i \,\Gamma^M \Gamma^N L_{MN} \Psi_1 +  \overline{\Psi}_2 i\, \Gamma^M \Gamma^N L_{MN} \Psi_2 \nonumber \\
&&  \hspace{1.5in} -\overline{\Phi}_1 i\, \Gamma^M \Gamma^N R_{MN} \Phi_1 -  \overline{\Phi}_2 i \,\Gamma^M \Gamma^N R_{MN} \Phi_2 \bigg] \, ,
\eeqa
where $\Gamma_M= (\gamma_\mu, \, i\gamma_5)$. The relative minus sign between $\Psi$ and $\Phi$ sectors is dictated by the parity invariance. Focusing on $M, N = \mu, 5$ terms, we have
\beqa
{\cal S}_{\rm dipole} &\supset& 2  M_5 R \int d^5 x \left( \frac{R}{z}\right)^3 \lambda_D \bigg[
-\overline{\Psi}_{1,L} \bar{\sigma}^\mu L_{\mu 5} \Psi_{1,L} +\overline{\Psi}_{1,R} \sigma^\mu L_{\mu 5} \Psi_{1,R} \nonumber \\
&& \hspace{1.7in} -\overline{\Psi}_{2,L} \bar{\sigma}^\mu L_{\mu 5} \Psi_{2,L} +\overline{\Psi}_{2,R} \sigma^\mu L_{\mu 5} \Psi_{2,R}  \nonumber \\
 && \hspace{1.7in}  +\overline{\Phi}_{1,L} \bar{\sigma}^\mu R_{\mu 5} \Phi_{1,L} -\overline{\Phi}_{1,R} \sigma^\mu R_{\mu 5} \Phi_{1,R} \nonumber \\
&& \hspace{1.7in} +\overline{\Phi}_{2,L} \bar{\sigma}^\mu R_{\mu 5} \Phi_{2,L} -\overline{\Phi}_{2,R} \sigma^\mu R_{\mu 5} \Phi_{2,R} \bigg]\nonumber  \\
&=&  2 M_5 R \int d^5 x \left( \frac{R}{z}\right)^3 \lambda_D\, \Tr \bigg[ - \overline{\cal H}_L \gamma_5 \gamma^\mu L_{\mu 5} {\cal H}_L + 
\overline{\cal H}_R \gamma_5 \gamma^\mu R_{\mu 5} {\cal H}_R \bigg] \nonumber \\
&=& \sqrt{2}   M_5 R \int d^5 x \left( \frac{R}{z}\right)^3 \lambda_D\, \Tr \bigg[ \overline{\cal H}^\prime  \gamma_5 \gamma^\mu V_{\mu 5} {\cal H}
+ \overline{\cal H}  \gamma_5 \gamma^\mu V_{\mu 5} {\cal H}^\prime \nonumber \\
&& \hspace{1.8in} +\overline{\cal H}^\prime  \gamma_5 \gamma^\mu A_{\mu 5} {\cal H}^\prime
+\overline{\cal H}  \gamma_5 \gamma^\mu A_{\mu 5} {\cal H} \bigg] \, ,
\label{eq:intra-multiplet-coupling}
\eeqa
where in the last step we have used Eqs.~(\ref{eq:bi-spinor-PD-H}), (\ref{eq:bi-spinor-PD-Hprime}), and (\ref{eq:gauge-field-relation}). The last term in Eq.~(\ref{eq:intra-multiplet-coupling}) gives rise to $D^* \to D\, \pi$ decay. In terms of 4D fields, the interaction can be written as
\beqa
{\cal L}_{\rm int}= -\frac{g_A}{F_\pi} \, \Tr \left[ \overline{H}(x) \gamma_5 \gamma^\mu \partial_\mu \boldsymbol \pi (x) H(x)\right] \, ,
\eeqa
with
\beqa
g_A =  -\sqrt{2} \,M_5 R\,F_\pi \,\int dz \left( \frac{z}{R}\right)^2 \lambda_D \left[f_{a,1}(z)\right]^2 f_0^\pi (z) \, .
\eeqa
It can be directly compared to the interaction written down in Ref.~\cite{Wise:1992hn,Burdman:1992gh,Yan:1992gz}. The coupling $g_A$ depends on the free parameter $\lambda_D$ which needs to be fit from the data. For $\nu=0$ and constant $\lambda_D$, $g_A=0.67 \lambda_D$. The measurements of $D^* \to D \pi$ decay width at CLEO and BaBar experiments correspond to $g_A= 0.59\pm0.01\pm0.07$~\cite{Anastassov:2001cw} and $0.570\pm 0.004\pm 0.005$~\cite{Lees:2013zna}, respectively.

\section{Discussion and Conclusions}
\label{sec:conclusions}

In this paper we discuss how to incorporate heavy-light mesons into AdS/QCD. The crucial observation is that in the heavy quark limit the heavy quark plays the role of a static color source and the dynamics is governed by the light quark. This prompts us to map the heavy-light mesons into fermions in the AdS bulk. The heavy quark spin symmetry can be incorporated as a global flavor symmetry in the bulk. We show that such a construction can give a reasonable description of the parity-doublets of the spin-0 and spin-1 heavy-light mesons. 
One may generalize it to heavy quark spin multiplets of meson states with higher angular momenta, or even baryons which contain both heavy and light quarks.
Of course this bottom-up AdS/QCD approach has its limited power because the real QCD is neither conformal nor close to the large $N_c$ limit. In our calculation, we have only considered the leading contributions in the heavy quark limit. The effects suppressed by inverse powers of the heavy quark mass are not included, such as the hyperfine splitting between the spin-0 and spin-1 mesons due to the interaction between the spins of the heavy quark and the light quark. Because we treat the heavy quark spin symmetry as a flavor symmetry in the bulk, such a mixing between the light quark spin and the bulk flavor symmetry violates the Lorentz symmetry in the AdS theory. It may be viewed as if the heavy quark provides an external chromomagnetic field due to its spin to the light quark in addition to the static color source. It would require additional parameters to fit these effects suppressed by the heavy quark masses.

We have only considered the strong interaction in this work. There are few predictions which may be checked against the limited amount of experimental and lattice data in this respect. Most of the spectrum and decay constant data serve to fix the model parameters. The real tests of the model can come when we consider weak interaction processes involving heavy-light mesons. As a first test, one can try to calculate the Isgur-Wise function~\cite{Isgur:1989vq,Isgur:1989ed} for the $B \to D$ transitions. In this case one needs to keep a finite velocity of the heavy quark and evaluate the overlap of the wave functions of the light quark in the initial and final states. We plan to study this in a future publication.

At low energies the weak interaction is described by dimension-six current interactions and we need to evaluate matrix elements of relevant current operators.  In the holographic model the weak currents can be generated by the corresponding sources on the UV boundary without additional parameters. In this way one could obtain ``predictions'' of various matrix elements in this model, which may provide some insights of the long-distance effects in the real QCD. For example, the LHCb experiment (and also the CDF experiment) earlier found large CP asymmetry $\Delta {\cal A}_{CP}\equiv {\cal A}_{CP}(D^0 \to K^+ K^-) - {\cal A}_{CP}(D^0 \to \pi^+ \pi^-)$ in $D$ meson decays~\cite{Aaij:2011in,Collaboration:2012qw}, much larger than many Standard Model estimations~\cite{Buccella:1994nf,Grossman:2006jg,Bigi:2011re,Isidori:2011qw,Cheng:2012wr,Li:2012cfa,Franco:2012ck}. Although the significance in the new LHCb data has greatly reduced~\cite{Aaij:2013bra}, a large $\Delta {\cal A}_{CP}$ is still possibly compatible with the data. It has been argued that a large $\Delta {\cal A}_{CP}$ may be accommodated if there is an enhancement of certain penguin matrix elements due to the long-distance effects in the Standard Model~\cite{Golden:1989qx,Isidori:2011qw,Brod:2011re,Pirtskhalava:2011va,Bhattacharya:2012ah,Feldmann:2012js,Brod:2012ud,Bhattacharya:2012kq}, analogous to the case of the ``$\Delta I=1/2$ rule'' in $K\to \pi\pi$ decay. For the $K\to \pi \pi$ decay, an AdS/QCD calculation indeed showed the enhancement of the matrix elements required to explain the $\Delta I=1/2$ rule~\cite{Hambye:2005up, Hambye:2006av}. With the AdS/QCD extended to heavy-light mesons one can carry out an analogous calculation which may shed lights on whether one can expect a large enhancement of the particular matrix elements due to long-distance effects. It will be left for future investigations.

\section*{Acknowledgment}
We would like to thank Hai-Yang Cheng, Tony Gherghetta, Hsiang-nan Li, Markus Luty, Ami Katz, Raman Sundrum and John Terning for useful discussion. H.-C.~Cheng would like to thank Fermilab Theory Group, Academia Sinica, and National Center for Theoretical Sciences (North) Physics Division in Taiwan for hospitality. We also would like to thank Aspen Center for Physics and Kvali Institute for Theoretical Physics, U.~C.~Santa Barbara, where part of this work was done. 
Y.~Bai is supported by startup funds from the UW-Madison. H.-C.~Cheng is supported in part by U.S. DOE grant No. DE-FG02-91ER40674. This research was also supported in part by the National Science Foundation under Grant No.\ NSF PHY11-25915 and Grant No.\ PHYS-1066293.

%---------------------------------------------------------------------
\appendix
%--------------------------------------------------------------------
%-----------------------------------------------------------------------------
\section{Weyl and Pauli-Dirac representations}
\label{sec:weyldirac}
%-----------------------------------------------------------------------------
We adopt the metric convention as $g_{\mu\nu}=g^{\mu\nu} = \mbox{diag}\{1, -1, -1, -1\}$. In the Weyl representation, the gamma matrices are
\beqa
\gamma^\mu = \left( 
\begin{array}{cc}
0  &  \sigma^\mu \\
\bar{\sigma}^\mu & 0 
\end{array}
\right) \,,
\qquad
\gamma^5 = \left( 
\begin{array}{cc}
-\mathbb{I}_2  &  0 \\
0 & \mathbb{I}_2
\end{array}
\right)\,,
\eeqa
with $\sigma^\mu = (1, \vec{\sigma})$ and $\bar{\sigma}^\mu = (1, -\vec{\sigma})$. In the Pauli-Dirac representation, the gamma matrices are
\beqa
\gamma^0 = \left( 
\begin{array}{cc}
\mathbb{I}_2  &  0 \\
0 & -\mathbb{I}_2 
\end{array}
\right) \,,
\qquad
\gamma^i = \left( 
\begin{array}{cc}
0 &  \sigma^i \\
-\sigma^i & 0
\end{array}
\right)\,,
\qquad
\gamma^5 = \left( 
\begin{array}{cc}
0 &  \mathbb{I}_2 \\
\mathbb{I}_2 & 0
\end{array}
\right)\,.
\eeqa
The transformation between Weyl and Pauli-Dirac representation is given by
\beqa
\left( \begin{array}{c} \psi_A \\ \psi_B \end{array} \right)_{\rm Weyl} = S 
\left( \begin{array}{c} \psi_A \\ \psi_B \end{array} \right)_{\rm PD}\,,\qquad \mbox{where}\,\quad\, S=\frac{1}{\sqrt{2}}
 \left( \begin{array}{cc} \mathbb{I}_2 &  -\mathbb{I}_2 \\ \mathbb{I}_2 & \mathbb{I}_2 \end{array} \right)\,.
\eeqa
For a bi-spinor representation, one can have independent representation for each spinor index. We keep the heavy spinor index in the Pauli-Dirac representation and only transform the light spinor index of a bi-spinor between the two representations, i.e.,
\beqa
{\cal H}^{\rm Weyl} = S\, {\cal H}^{\rm PD}\,, \qquad 
{\cal H}^{\rm PD} = S^{-1}\, {\cal H}^{\rm Weyl} \,.
\eeqa
%

%%%%%%%%%%%%%%%%%%%%%%%%%%%%%%%%%%%
%\bibliography{AdsHQET}
%\bibliographystyle{JHEP}
\providecommand{\href}[2]{#2}\begingroup\raggedright\endgroup

\end{document}